\documentclass{webofc}
\usepackage[varg]{txfonts}   
\usepackage{listings,wrapfig}
\usepackage{hyperref}
\begin{document}
\title{Toward a renewed Galactic Cepheid distance scale from \emph{Gaia} and optical interferometry}
%
%

\author{\firstname{Pierre} \lastname{Kervella}\inst{1,2}\fnsep\thanks{\href{mailto:pierre.kervella@obspm.fr}{\tt pierre.kervella@obspm.fr}} \and
        \firstname{Antoine} \lastname{M\'erand}\inst{3} \and
        \firstname{Alexandre} \lastname{Gallenne}\inst{3} \and
        \firstname{Boris} \lastname{Trahin}\inst{1,2} \and
        \firstname{Nicolas} \lastname{Nardetto}\inst{4} \and
        \firstname{Richard I.} \lastname{Anderson}\inst{5} \and
        \firstname{Joanne} \lastname{Breitfelder}\inst{2,3} \and
        \firstname{Laszlo} \lastname{Szabados}\inst{6} \and
	\firstname{Howard E.} \lastname{Bond}\inst{7} \and
        \firstname{Simon} \lastname{Borgniet}\inst{1,2} \and
        \firstname{Wolfgang} \lastname{Gieren}\inst{8} \and
        \firstname{Grzegorz} \lastname{Pietrzy{\'n}ski}\inst{9}
}

\institute{
Unidad Mixta Internacional Franco-Chilena de Astronom\'{i}a (CNRS UMI 3386), Departamento de Astronom\'{i}a, Universidad de Chile, Las Condes, Chile.
\and
LESIA (CNRS UMR 8109), Observatoire de Paris, PSL, CNRS, UPMC, Univ. Paris-Diderot, France.
\and
European Southern Observatory, Alonso de C\'ordova 3107, Casilla 19001, Santiago 19, Chile.
\and
Laboratoire Lagrange, UMR 7293, Univ. de Nice Sophia-Antipolis, CNRS, Obs. de la C{\^o}te d'Azur, France
\and
Physics and Astronomy Department, The Johns Hopkins University, Baltimore, MD 21218, USA
\and
Konkoly Observatory, MTA CSFK, Konkoly Thege M. {\'u}t 15-17, H-1121, Hungary
\and
Department of Astronomy \& Astrophysics, Pennsylvania State University, University Park, PA 16802 USA.
\and
Universidad de Concepci{\'o}n, Departamento de Astronom\'{\i}a, Casilla 160-C, Concepci{\'o}n, Chile.
\and
Nicolaus Copernicus Astronomical Center, Polish Academy of Sciences, Warszawa, Poland.
          }

\abstract{%
   Through an innovative combination of multiple observing techniques and modeling, we are assembling a comprehensive understanding of the pulsation and close environment of Cepheids.
   We developed the SPIPS modeling tool that combines all observables (radial velocimetry, photometry, angular diameters from interferometry) to derive the relevant physical parameters of the star (effective temperature, infrared excess, reddening,...) and the ratio of the distance and the projection factor $d/p$.
   We present the application of SPIPS to the long-period Cepheid RS Pup, for which we derive $p = 1.25 \pm 0.06$.
   The addition of this massive Cepheid consolidates the existing sample of $p$-factor measurements towards long-period pulsators.
   This allows us to conclude that $p$ is constant or mildly variable around $p = 1.29 \pm 0.04$ ($\pm 3\%$) as a function of the pulsation period.
   The forthcoming \emph{Gaia} DR2 will provide a considerable improvement in quantity and accuracy of the trigonometric parallaxes of Cepheids.
   From this sample, the SPIPS modeling tool will enable a robust calibration of the Cepheid distance scale.
}
\maketitle

\section{Introduction}\label{sec:intro}

The empirical Leavitt law of Cepheids (the Period-Luminosity relation) \cite{1912HarCi.173....1L} is a cornerstone of the observational calibration of the cosmological distance scale.
The persistent tension between the determinations of the Hubble constant $H_0$ from the CMB modeling and the empirical distance ladder may have far-reaching implications on, for instance, the equation of state of dark energy.
The calibration of the empirical ladder relies primarily on Cepheids and Type Ia supernovae (\cite{2016Riess}, see also \cite{2017Bonvin}), and an improvement of the true calibration accuracy of the Leavitt law is required to confirm the significance of the tension.
%
The parallax-of-pulsation method (PoP), also known as the Baade-Wesselink technique, is a powerful way to measure the distances to individual Galactic and Magellanic Cloud Cepheids, and thus calibrate the Leavitt law.
It relies on the comparison of the amplitude of the pulsation of the star from
(1) the integration of its radial velocity curve measured using spectroscopy and
(2) the change in angular diameter of the star estimated from its brightness and color, or measured using interferometry.
The major weakness of the PoP technique is that it uses a numerical factor to convert disk-integrated radial velocities into photospheric velocities: the projection factor, or $p$-factor (\cite{2007Nardetto, 2009Barnes, 2017Nardetto}).
The PoP technique provides the ratio $d/p$, but the distance $d$ and the $p$-factor are fully degenerate.
The calibration of the $p$-factor of Cepheids therefore requires to know $d$ independently.
At the moment, only a dozen reasonably accurate trigonometric parallaxes of nearby Cepheids is available, mainly from HST's Fine Guidance Sensor (\cite{2002Benedict,2007Benedict}) and WFC3 (\cite{2016Casertano}).
But \emph{Gaia} will soon resolve the $d/p$ degeneracy for hundreds of Cepheids, thus removing the major systematics of the PoP technique.
The outcome will be a robust calibration of the Cepheid distance scale, based on well-known Galactic and Magellanic Cloud pulsators.
For this program, we developed an innovative version of the PoP technique, the {\it Spectro-Photo-Interferometry of Pulsating Stars} (SPIPS, Sect.~\ref{spips}) that simultaneous integrates multiple observables in a consistent model of the pulsation of the star.
We present in Sect.~\ref{spips-cepheids} its application to the Cepheid RS\,Pup and the dependence of the $p$-factor on the pulsation period. We discuss in Sect.~\ref{gaia} the improvement that will come from \emph{Gaia}'s DR2 parallaxes in 2018.

\section{The SPIPS algorithm \label{spips}}

The general principle of the SPIPS algorithm \cite{2015Merand} is to fit a parametric model of the star's pulsation simultaneously to the full set of observables: radial velocities from spectroscopy, photometry in all photometric bands and (when available) angular diameter measurements from optical interferometry.

\begin{wrapfigure}[18]{r}{0.45\textwidth}
\includegraphics[width=0.45\textwidth]{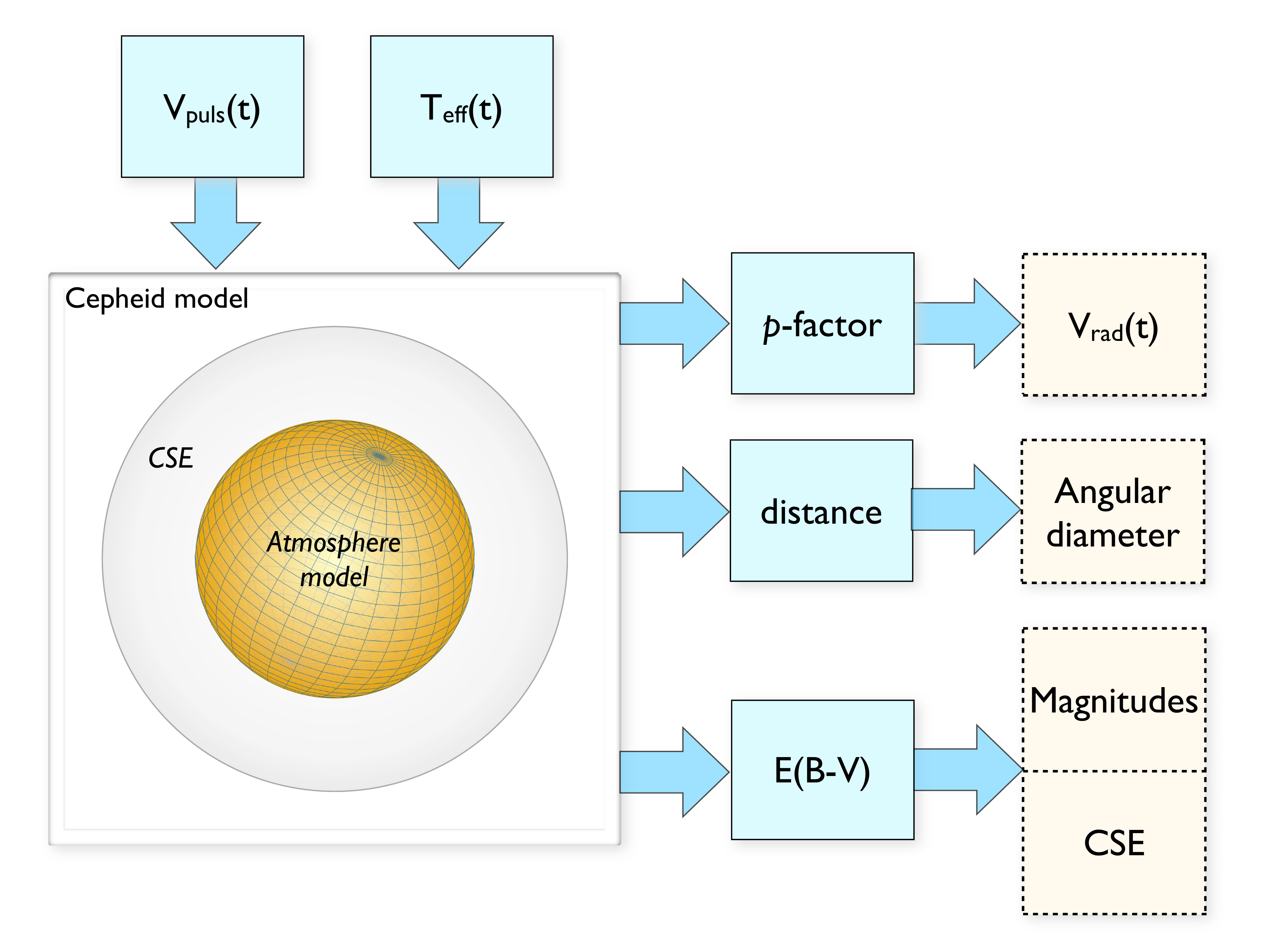}
\caption{Overview of the principle of the SPIPS modeling. The main model parameters are shown on a light blue background, and the  produced observables are shown on the right.}\label{wrap-fig:1}
\end{wrapfigure} 

The SPIPS model of the star is built assuming that Cepheids are radially pulsating spheres, for which the pulsational velocity $v_\mathrm{puls}(t)$ and the effective temperature $T_\mathrm{eff}(t)$ are the basic descriptive parameters (Fig.~\ref{wrap-fig:1}).
The cyclic variation of these parameters is represented using classical Fourier series or periodic splines functions.
The photometry in all filters is computed using ATLAS9\footnote{\tt\url{http://wwwuser.oats.inaf.it/castelli/grids.html}} atmosphere models, considering the bandpass and zero points matching the observations (using the SVO\footnote{\tt\url{http://svo2.cab.inta- csic.es/}} or Asiago\footnote{\tt\url{http://ulisse.pd.astro.it/Astro/ADPS/Paper/index. html}} databases).
The interstellar reddening is parametrized using the standard $E(B-V)$ color excess for typical Milky Way dust ($R_V=3.1$) and computed for the phase-variable $T_\mathrm{eff}$ of the Cepheid.
Angular diameters including limb darkening (LD) are produced by the model to match the interferometric measurements, using SATLAS models (\cite{2013Neilson}).
Finally, circumstellar envelopes are included in the modeling, considering both their photometric contributions in the infrared $H$ and $K$ bands and their influence on interferometric angular diameter measurements.

\section{The projection factor of Cepheids\label{spips-cepheids}}
%
\begin{wrapfigure}[17]{r}{5.2cm}
\includegraphics[width=5.2cm]{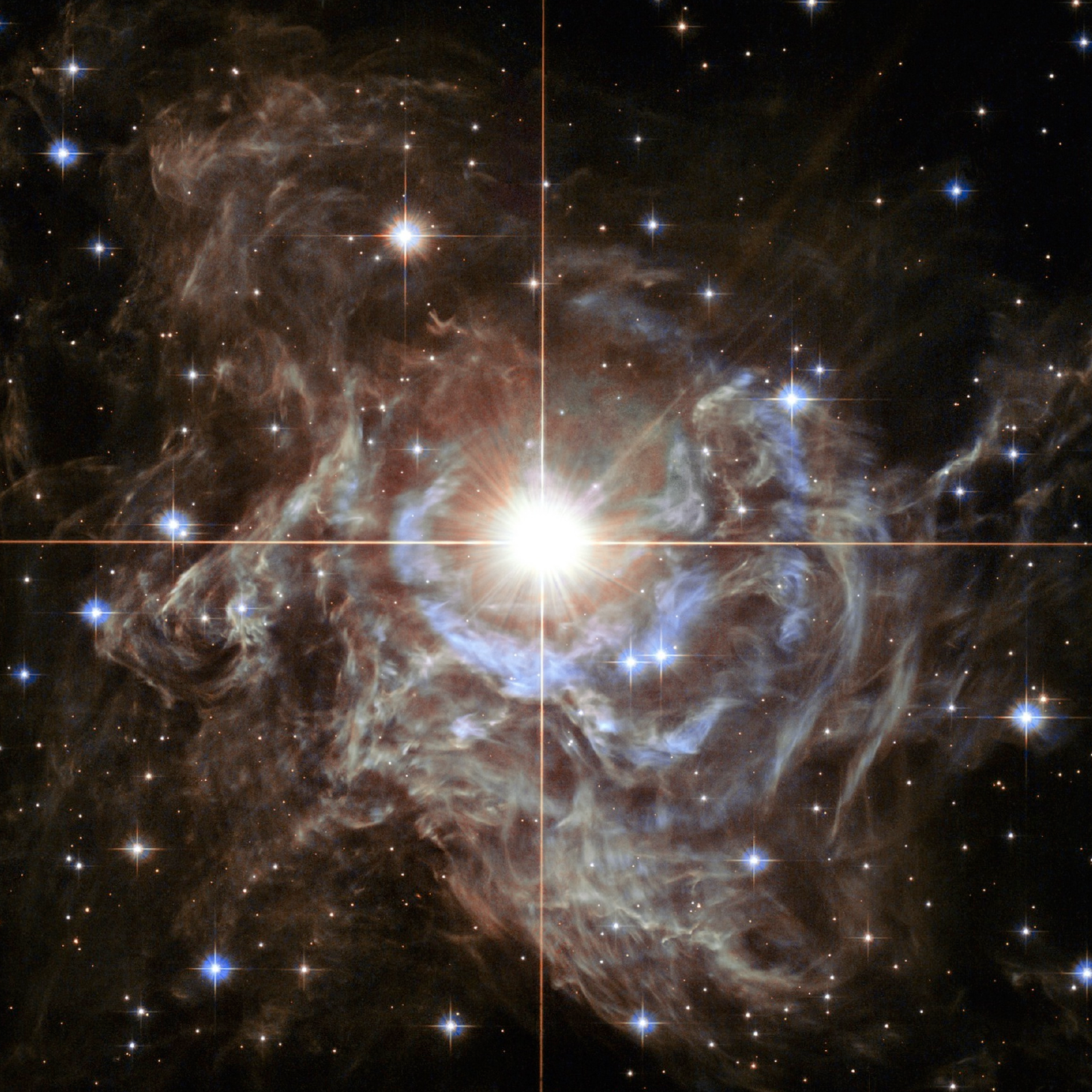}
\caption{Color composite image of RS\,Pup and its nebula (from \cite{2014Kervella}).}\label{wrap-fig:2}
\end{wrapfigure} 

Recent applications of the SPIPS parametric modeling to bright Galactic Cepheids are presented in \cite{2015Merand} ($\delta$\,Cep, $\eta$\,Aql), \cite{2016Breitfelder} (nine Cepheids with HST/FGS parallaxes) and \cite{2015Breitfelder} (Type 2 Cepheid $\kappa$\,Pav).

\begin{figure*}
\centering
\includegraphics[width=\hsize,clip]{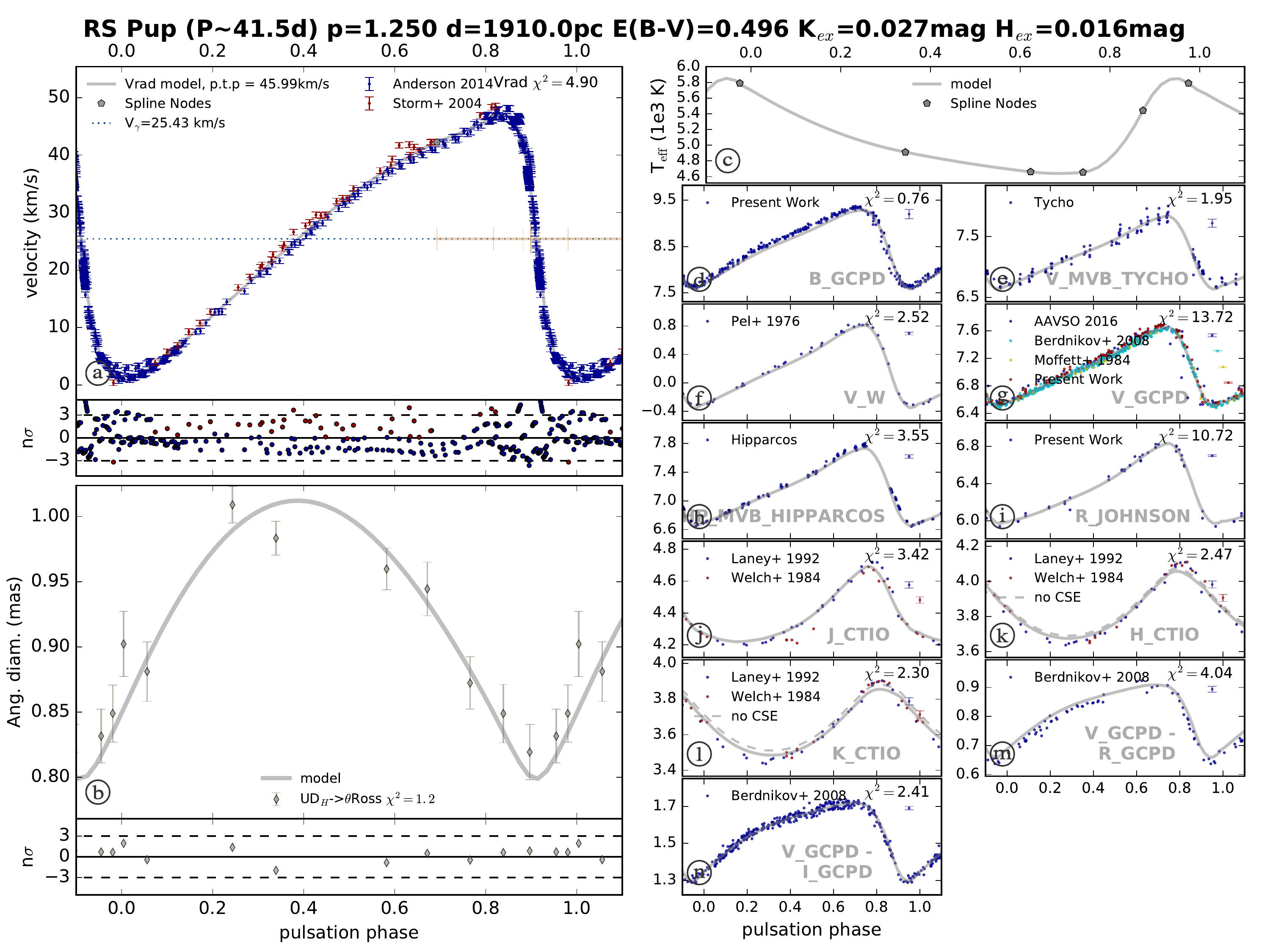}
\caption{SPIPS combined fit of the available observations of RS\,Pup (figure from \cite{2017Kervella}). The radial velocity measurements are presented in the upper left panel, the interferometric angular diameters in the lower left panel, and the different photometric bands in the sub-panels on the right. The effective temperature of the model is shown in the upper right panel. The best-fit model curves are represented using light grey solid lines.}
\label{fig:fig-3}       
\end{figure*}

A SPIPS model of the long-period Cepheid RS\,Pup ($P=41.5$\,days) has recently been presented by \cite{2017Kervella}. 
RS\,Pup is one of the intrinsically brightest Cepheids in the Galaxy, and it is particularly remarkable due to its large circumstellar nebula (Fig.~\ref{wrap-fig:2}) that reflects the light variations of the Cepheid \cite{2008Kervella, 2009Kervella, 2012Kervella, 2012KervellaMessenger, 2014Kervella, 2017Kervella}.
From a combination of photometry and polarimetry of the light echoes in the nebula, \cite{2014Kervella} obtained a distance of $d=1910 \pm 80$\,pc ($\pm 4.2\%$), corresponding to a parallax $\pi = 0.524 \pm 0.022$\,mas.
Knowing the distance of RS\,Pup opens the possibility to resolve the distance/$p$-factor degeneracy and determine the value of $p$.
An overview of the SPIPS model of RS\,Pup is presented in Fig.~\ref{fig:fig-3}.
It was computed using as constraints the photometric and radial velocity measurements available in the literature (e.g., from \cite{2014Anderson}), completed with a series of VLTI/PIONIER angular diameter measurements (\cite{2017Kervella}).
Interferometric angular diameters are not mandatory for the SPIPS model to converge.
But as they are insensitive to reddening, they allow to efficiently constrain both the effective temperature $T_\mathrm{eff}$ and the color excess $E(B-V)$ of the star, that are otherwise correlated.
The overall quality of the fit of the available observables is very good, with a best-fit $p$-factor of $p = 1.25 \pm 0.06$ ($\pm 5\%$).
The color excess $E(B-V)$ is estimated to $0.496 \pm 0.006$ and a moderate infrared excess of 0.02 to 0.03\,mag ($\pm 0.01$) is found in the $HK$ bands.

\begin{figure}
\centering
\sidecaption
\includegraphics[width=8cm,clip]{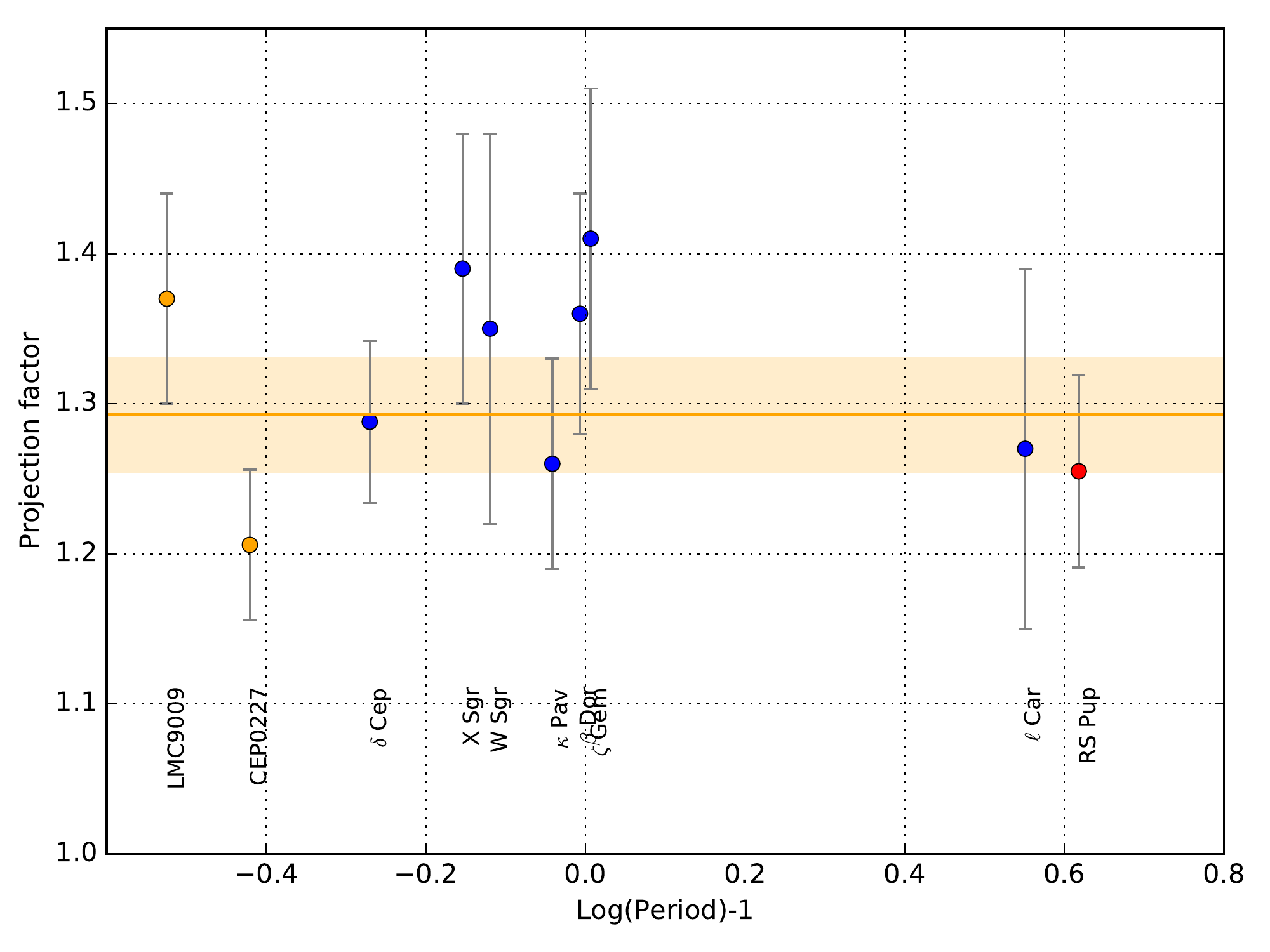}
\caption{Measured p-factors of Cepheids with better than 10\% relative accuracy (figure from \cite{2017Kervella}). The blue points use HST-FGS distances (\cite{2002Benedict,2007Benedict}), the orange points are the LMC eclipsing Cepheids (\cite{{2013Pilecki},2015Gieren}) and the red point is RS\,Pup.
The solid line and orange shaded area represent the weighted average of the independent measurements ($p = 1.29 \pm 0.04$).}
\label{fig:fig-4}       
\end{figure}

The addition of the $p$-factor of RS\,Pup to the limited set of existing $p$-factor measurements is particularly valuable, as the sample was up to now mostly limited to short and intermediate period Cepheids (Fig.~\ref{fig:fig-4}).
The other exception is $\ell$\,Car ($P=35.5$\,days), whose parallax was measured by \cite{2007Benedict}, and whose $p$-factor was derived by \cite{2016Breitfelder} and \cite{2016Anderson}.
The $p$-factors of $\ell$\,Car and RS\,Pup are statistically identical within their uncertainties.
The simple model of a constant $p = 1.29 \pm 0.04$ over the sampled range of Cepheid periods (3.0 to 41.5\,days) is consistent with the observations ($\chi^2_\mathrm{red}=0.9$, \cite{2017Kervella}).

\section{Perspectives with \emph{Gaia} parallaxes\label{gaia}}

The precision of the TGAS parallaxes of Galactic Cepheids released in the \emph{Gaia} DR1 (\cite{2016Lindegren}) is insufficient to constrain significantly their $p$-factors (see e.g., \cite{2017Casertano}).
The \emph{Gaia} DR2 (currently expected in April 2018) will considerably improve the accuracy of Galactic Cepheid parallaxes.
Based on the foreseen accuracy of \emph{Gaia}, 400+ Cepheid parallaxes will be measured with an accuracy better than 3\%, out of which approximately 100+ with an accuracy better than 1\%.
The ongoing observations of a sample of 18 long-period Cepheid using the spatial scanning mode of the HST/WFC3 by \cite{2016Casertano} will also provide a complementary set of parallaxes for these rare pulsators.
The flexibility of the SPIPS modeling enables the integration of all types of new and archival observations in a consistent and robust approach, thus mitigating the influence of systematics of observational origin.
Taking into account the contribution of Cepheid envelopes, that can be very significant at infrared wavelengths (\cite{2012Gallenne,2013Gallenne}), as well as the presence of stellar companions (\cite{2015Gallenne, 2017Gallenne}) will also contribute to mitigate the astrophysical biases on the calibration of the Leavitt law.

\begin{acknowledgement} 
\noindent\vskip 0.2cm
\noindent {\em Acknowledgments}: The authors acknowledge the support of the French Agence Nationale de la Recherche (ANR), under grant ANR-15-CE31-0012-01 (project UnlockCepheids).
PK, AG, and WG acknowledge support of the French-Chilean exchange program ECOS-Sud/CONICYT (C13U01).
W.G. and G.P. gratefully acknowledge financial support for this work from the BASAL Centro de Astrofisica y Tecnologias Afines (CATA) PFB-06/2007. 
W.G. also acknowledges financial support from the Millenium Institute of Astrophysics (MAS) of the Iniciativa Cientifica Milenio del Ministerio de Economia, Fomento y Turismo de Chile, project IC120009.
The research leading to these results  has received funding from the European Research Council (ERC) under the European Union's Horizon 2020 research and innovation programme (grant agreement No 695099).
HEB acknowledges support from NASA through grant GO-11715 from the Space Telescope Science Institute, which is operated by the Association of Universities for Research in Astronomy, Inc., under NASA contract NAS5-26555

\end{acknowledgement}

%
%

\end{document}